\documentclass[preprintnumbers,10pt,nofootinbib]{revtex4}

\usepackage{amsmath,latexsym,amssymb,amsfonts}
\usepackage[dvips]{graphicx,color}
\usepackage{bm}

\begin{document}


\title{\textbf{Tunneling from the past horizon}}

\author{
\textsc{Subeom Kang$^{a,b}$}\footnote{{\tt ksb527{}@{}kaist.ac.kr}} and \textsc{Dong-han Yeom$^{c,d}$}\footnote{{\tt innocent.yeom{}@{}gmail.com}}
}

\affiliation{
$^{a}$Department of Physics, KAIST, Daejeon 34141, Republic of Korea\\
$^{b}$Center for Theoretical Physics of the Universe, Institute for Basic Science (IBS), Daejeon 34051, Republic of Korea\\
$^{c}$Asia Pacific Center for Theoretical Physics, Pohang 37673, Republic of Korea\\
$^{d}$Department of Physics, POSTECH, Pohang 37673, Republic of Korea
}

\begin{abstract}
We investigate a tunneling and emission process of a thin-shell from a Schwarzschild black hole, where the shell was initially located beyond the Einstein-Rosen bridge and finally appears at the right side of the Penrose diagram. In order to obtain such a solution, we should assume that the areal radius of the black hole horizon increases after the tunneling. Hence, there is a parameter range such that the tunneling rate is exponentially enhanced, rather than suppressed. We may have two interpretations regarding this. First, such a tunneling process from the past horizon is improbable by physical reasons; second, such a tunneling is possible in principle, but in order to obtain a stable Einstein-Rosen bridge, one needs to restrict the parameter spaces. If such a process is allowed, this can be a non-perturbative contribution to Einstein-Rosen bridges as well as eternal black holes.
\end{abstract}

\maketitle

\newpage

\tableofcontents


\section{Introduction}

A black hole is a classical object, but it emits particles via quantum processes. The most famous example is known by Hawking radiation \cite{Hawking:1974sw}. There are several points of view to understand this. One of the most interesting approaches toward Hawking radiation is to interpret the radiation as a particle tunneling process from the black hole \cite{Hartle:1976tp,Parikh:1999mf,Srinivasan:1998ty}.

In the particle tunneling picture, one considers a particle that is initially inside the horizon. The particle moves backward in time, locates outside the horizon, and finally moves forward in time to the future infinity \cite{Hartle:1976tp}. This is not allowed by a classical way, but it can be possible quantum mechanically. The quantum process is described by the wave function, while the wave function can be well approximated by the Euclidean on-shell trajectory or the WKB approximation. Finally, we can obtain the Boltzmann factor as the emission rate or the decay rate, which is consistent with the result based on the Bogoliubov transformation \cite{Hawking:1974sw}.

Recently, there is an interesting observation such that the black hole is not only a site of Hawking radiation, but also a site of a bubble nucleation \cite{Gregory:2013hja}. A black hole can enhance the vacuum decay process, where the black hole can even be disappeared in some parameters \cite{Chen:2017suz}. These properties have physical implications for cosmology \cite{Farhi:1989yr} as well as the information loss problem of black holes \cite{Chen:2014jwq,Sasaki:2014spa}.

In this context, almost all investigations have been focused on the tunneling from the \textit{future} event horizon. On the other hand, what happens if there exists a past horizon? The traditional interpretation about the past horizon is related to the white hole, where this is regarded as unphysical and must be hidden by the star interior for realistic situations. However, in order to simplify mathematical descriptions of quantum processes, it is allowed to mathematically extend the solution and use the white hole region, where this was used in the original derivation of Hartle and Hawking \cite{Hartle:1976tp}. In addition to this, recently Maldacena and Susskind suggested an idea about the correspondence between the Einstein-Rosen (ER) bridge and the Einstein-Podolsky-Rosen (EPR) entanglements, so-called the ER=EPR conjecture \cite{Maldacena:2013xja}. In this idea, the causal structure beyond the Einstein-Rosen bridge as well as the causal structure near the white hole region become important in order to reconcile the problem of black hole complementarity \cite{Susskind:1993if} and the firewall \cite{Almheiri:2012rt}.

Following the ER=EPR conjecture, if we regard that the other side of the Einstein-Rosen bridge is not only a mathematical extension but also a physical reality of the world, then we need to care about the possible quantum effects across the Einstein-Rosen bridge. In order to do this, we consider a thin-shell dynamics \cite{Israel:1966rt,Chen:2017pkl}, where outside is a Schwarzschild solution and inside is a Schwarzschild-anti-de Sitter (AdS) solution. One can consider a situation that initially the shell is in the left side of the ER bridge and the shell tunnels to the right side of the Einstein-Rosen bridge. Hence, one can interpret that such a process is a \textit{tunneling from the past horizon}.

Perhaps one may choose a good quantum state and may avoid unwanted physical consequences. However, at once any strange quantum effects are allowed within the semi-classical description, we need to explain the reason why there is such an effect or why such an event will not happen in realistic situations. In this paper, we discuss various conceptual problems regarding this quantum process.

This paper is organized as follows. In Sec.~\ref{sec:thinshell}, we introduce a thin-shell model that allows a tunneling from inside to outside the past horizon. In Sec.~\ref{sec:prob}, we discuss the details of the tunneling rate and comment about possible problems and interesting applications. Finally, in Sec.~\ref{sec:con}, we summarize our discussion and give several comments on future research topics.

\section{\label{sec:thinshell}Thin-shell model}

In this section, we briefly discuss on the thin-shell model for a tunneling process, where the shell starts from the other side of the Einstein-Rosen bridge.

We consider a spacetime with the spherical symmetric metric ansatz
\begin{eqnarray}
\label{eq:metric}
ds_{\pm}^{2}= - f_{\pm}(R) dT^{2} + \frac{1}{f_{\pm}(R)} dR^{2} + R^{2} d\Omega^{2}.
\end{eqnarray}
In addition, we prepare a thin-shell that locates at $r$: outside the shell is $r < R$ (denoted by $+$) and inside the shell is $R < r$ (denoted by $-$). The thin-shell is defined on the induced metric
\begin{eqnarray}
ds^{2} = - dt^{2} + r^{2}(t) d\Omega^{2}.
\end{eqnarray}

We impose the metric ansatz for outside and inside the shell as follows:
\begin{eqnarray}
f_{\pm}(R) = 1 - \frac{2M_{\pm}}{R} + \frac{R^{2}}{\ell_{\pm}^{2}}.
\end{eqnarray}
Here, $M_{+}$ and $M_{-}$ are the mass parameters of each region and
\begin{eqnarray}
\ell^{2}_{\pm} = \frac{3}{8\pi |V_{\pm}|}
\end{eqnarray}
is the parameter due to the vacuum energy $V_{\pm}$. We especially assume $M_{-} = M_{+} + \omega$ with positive $\omega$ and $\ell_{+} = \infty$ (hence, outside is Schwarzschild, but we can easily generalize to the asymptotic anti-de Sitter).

One can derive the equation of motion of the thin-shell following the Israel junction equation \cite{Israel:1966rt}:
\begin{eqnarray}\label{eq:junc}
\epsilon_{-} \sqrt{\dot{r}^{2}+f_{-}(r)} - \epsilon_{+} \sqrt{\dot{r}^{2}+f_{+}(r)} = 4\pi r \sigma,
\end{eqnarray}
where $\epsilon_{\pm} = \pm 1$ shows the signs of the extrinsic curvatures and $\sigma$ is the tension parameter of the shell. We choose positive $\sigma$ so that this satisfies the null energy condition. After several straightforward calculations (see Appendix A), we simplify the junction equation as follows \cite{Blau:1986cw}:
\begin{eqnarray}\label{eq:form}
\dot{r}^{2} + V(r) &=& 0,\\\label{eq:form2}
V(r) &=& f_{+}(r)- \frac{\left(f_{-}(r)-f_{+}(r)-16\pi^{2} \sigma^{2} r^{2}\right)^{2}}{64 \pi^{2} \sigma^{2} r^{2}}\\
&=& 1 - \left(-\frac{1}{\ell_{+}^{2}} + \frac{\mathcal{B}^{2}}{64 \pi^{2} \sigma^{2}}\right) r^{2} - 2M_{+}\left( 1 - \frac{\omega}{M_{+}} \frac{\mathcal{B}}{32 \pi^{2} \sigma^{2}} \right) \frac{1}{r}  - \frac{\omega^{2}}{16 \pi^{2} \sigma^{2}} \frac{1}{r^{4}},
\end{eqnarray}
where
\begin{eqnarray}
\mathcal{B} \equiv \frac{1}{\ell_{-}^{2}} - \frac{1}{\ell_{+}^{2}} - 16 \pi^{2} \sigma^{2}.
\end{eqnarray}

The extrinsic curvatures are \cite{Blau:1986cw}
\begin{eqnarray}
\beta_{\pm}(r) \equiv \frac{f_{-}(r)-f_{+}(r)\mp 16\pi^{2} \sigma^{2} r^{2}}{8 \pi \sigma r} = \epsilon_{\pm} \sqrt{\dot{r}^{2}+f_{\pm}(r)}.
\end{eqnarray}
Therefore,
\begin{eqnarray}\label{eq:beta}
\beta_{\pm}(r) \propto - \frac{2 \omega}{r} + \left(\frac{1}{\ell_{-}^{2}} \mp 16 \pi^{2} \sigma^{2}\right) r^{2}.
\end{eqnarray}
This means that as $r$ goes to zero, $\epsilon_{\pm}$ are always $-1$. As $r$ goes to infinity, if $\ell_{-}^{-2} - 16\pi^{2} \sigma^{2} > 0$, then $\epsilon_{\pm}$ are always $+1$. In addition, by choosing a suitable $\omega$, around the turning point ($\dot{r} = 0$), it is possible to choose $\epsilon_{\pm} = -1$.

\begin{figure}
\begin{center}
\includegraphics[scale=1]{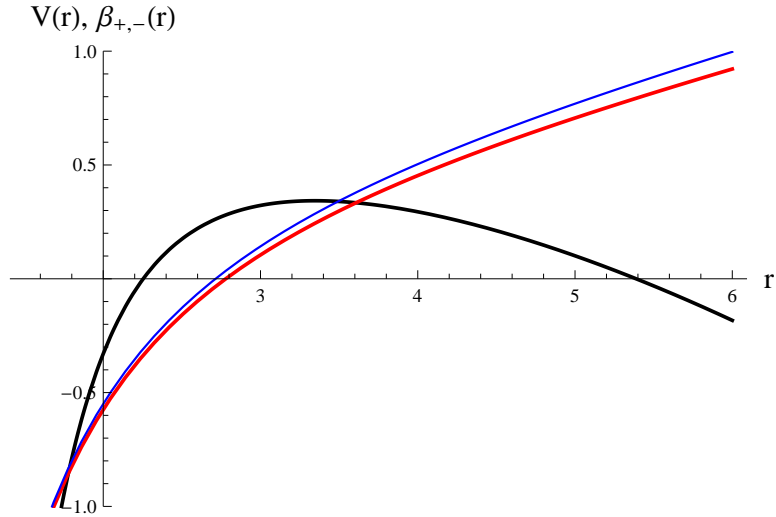}
\caption{\label{fig:example}An example of $V(r)$ (black), $\beta_{+}(r)$ (red), and $\beta_{-}$ (blue), where $M_{+} = 1$, $M_{-} = 1.046$, $\ell_{+} = \infty$, $\ell_{-} = 15$, and $\sigma = 0.001$.}
\end{center}
\end{figure}

\begin{figure}
\begin{center}
\includegraphics[scale=0.6]{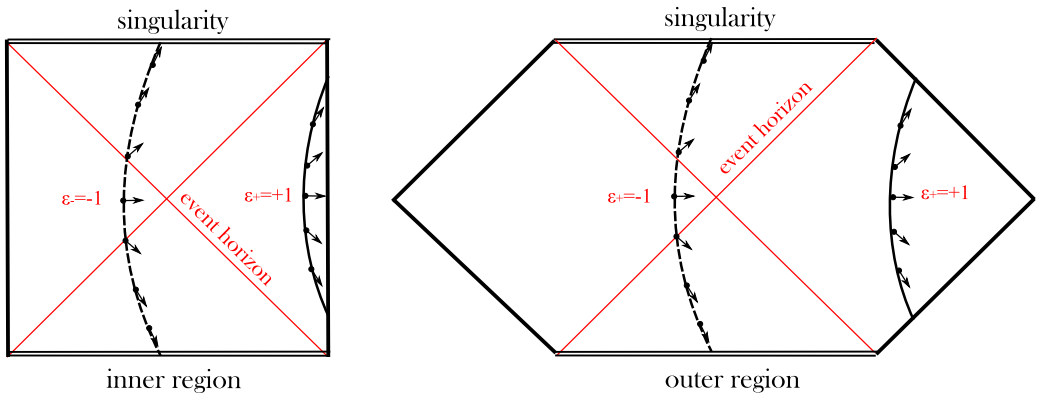}
\caption{\label{fig:signs}Dashed curves denote a shell trajectory of a symmetric collapsing solution and thick curves denote a shell trajectory of a symmetric bouncing solution; left is for the inside and right is for the outside the shell.}
\end{center}
\end{figure}

In this paper, we are interested in the case such that $V(r)$ has two zeros, say $r_{1} < r_{2}$ and $V(r_{1,2}) = 0$. Fig.~\ref{fig:example} is such an example of $V(r)$ as well as corresponding $\beta_{\pm}$. For given parameters, the effective potential $V(r)$ has two turning points. Since the effective potential has two turning points, there are two types of classical trajectories, where one is a symmetric collapsing solution (starts from zero, reaches $r_{1}$, and collapses to zero again) and the other is a symmetric bouncing solution (starts from infinity, reaches $r_{2}$, and expands to infinity again). Note that if $r \leq r_{1}$, both of $\beta_{\pm}$ are negative, while if $r \geq r_{2}$, both of $\beta_{\pm}$ are positive. Hence, if the shell trajectory is a symmetric collapsing solution, then the extrinsic curvature is always negative and hence the shell should be located beyond the Einstein-Rosen bridge (dashed curves of Fig.~\ref{fig:signs}). If the shell trajectory is a symmetric bouncing solution, then the shell should be located in the right side of the Penrose diagram (thick curves of Fig.~\ref{fig:signs}). In the next section, we will analyze more detailed parameter spaces that satisfy these conditions.

\begin{figure}
\begin{center}
\includegraphics[scale=0.6]{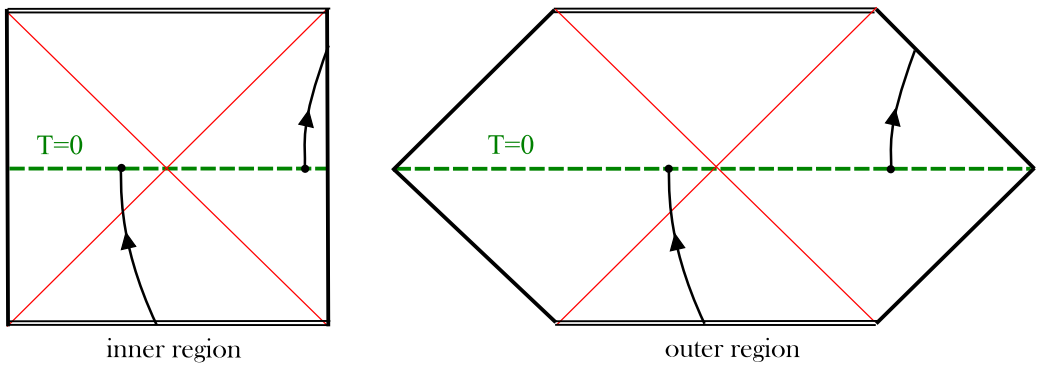}
\caption{\label{fig:tunneling}Tunneling of a shell from inside to outside the past horizon.}
\end{center}
\end{figure}

\section{\label{sec:prob}Probability and related issues}

In this section, we describe the tunneling probability of the shell from the symmetric collapsing shell to the symmetric bouncing shell.

\subsection{Tunneling rate}

Now we can consider a tunneling process between two turning points $r_{1}$ and $r_{2}$, i.e., a tunneling from the collapsing to the bouncing shell (Fig.~\ref{fig:tunneling}). Then this can be interpreted as a tunneling from inside the past horizon to outside the past horizon.

The tunneling probability can be obtained by introducing the Wick-rotation to the Euclidean time $\tau = it$ and by integrating the Euclidean action $S_{\mathrm{E}}$ of the shell. The decay rate $\Gamma$ is
\begin{eqnarray}
\Gamma \propto e^{-2B}
\end{eqnarray}
and following the formula of Gregory, Moss and Withers \cite{Gregory:2013hja} (that is consistent with \cite{Farhi:1989yr,Chen:2015ibc}, see Appendix B), we eventually obtain
\begin{eqnarray}\label{eq:formula}
2 B = - \Delta \left( \frac{\mathcal{A}_{h}}{4}\right) + \frac{1}{4} \int d\tau \left[ \left( 2 r f_{+} - r^{2} f_{+}' \right) \dot{T}_{+} - \left(  2 r f_{-} - r^{2} f_{-}' \right) \dot{T}_{-} \right].
\end{eqnarray}
Here, $\Delta \mathcal{A}_{h}/4 \equiv (\mathcal{A}_{-} - \mathcal{A}_{+})/4$, where $\mathcal{A}_{+}$ and $\mathcal{A}_{-}$ are the area of the horizon for outside and inside the shell, respectively. $f'_{\pm}$ is the derivation with respect to $r$ and $\dot{T}_{\pm}$ is the derivation with respect to $\tau$.

One important observation is that the first term $- \Delta \mathcal{A}_{h}/4$ is negative definite because of positive $\omega$. This causes a possibility that $2B < 0$, or such a process is exponentially enhanced.

\begin{figure}
\begin{center}
\includegraphics[scale=0.54]{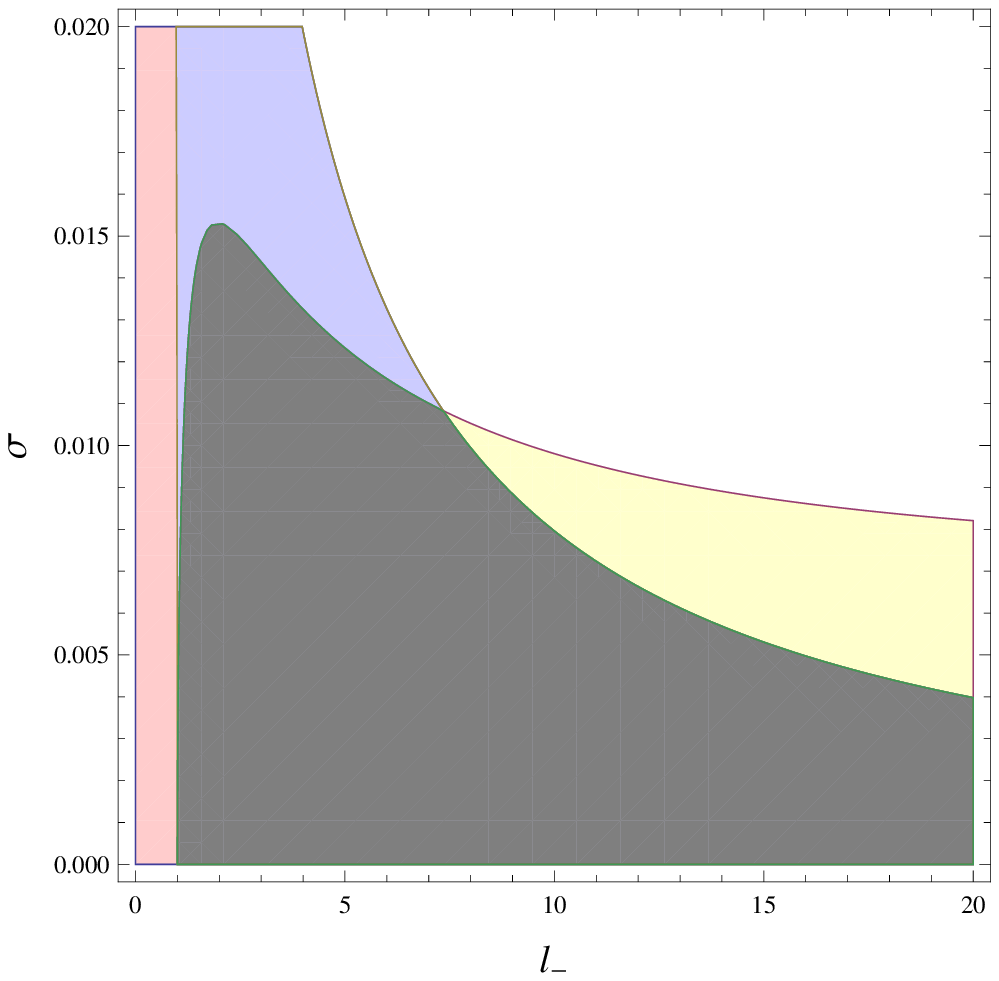}
\includegraphics[scale=0.54]{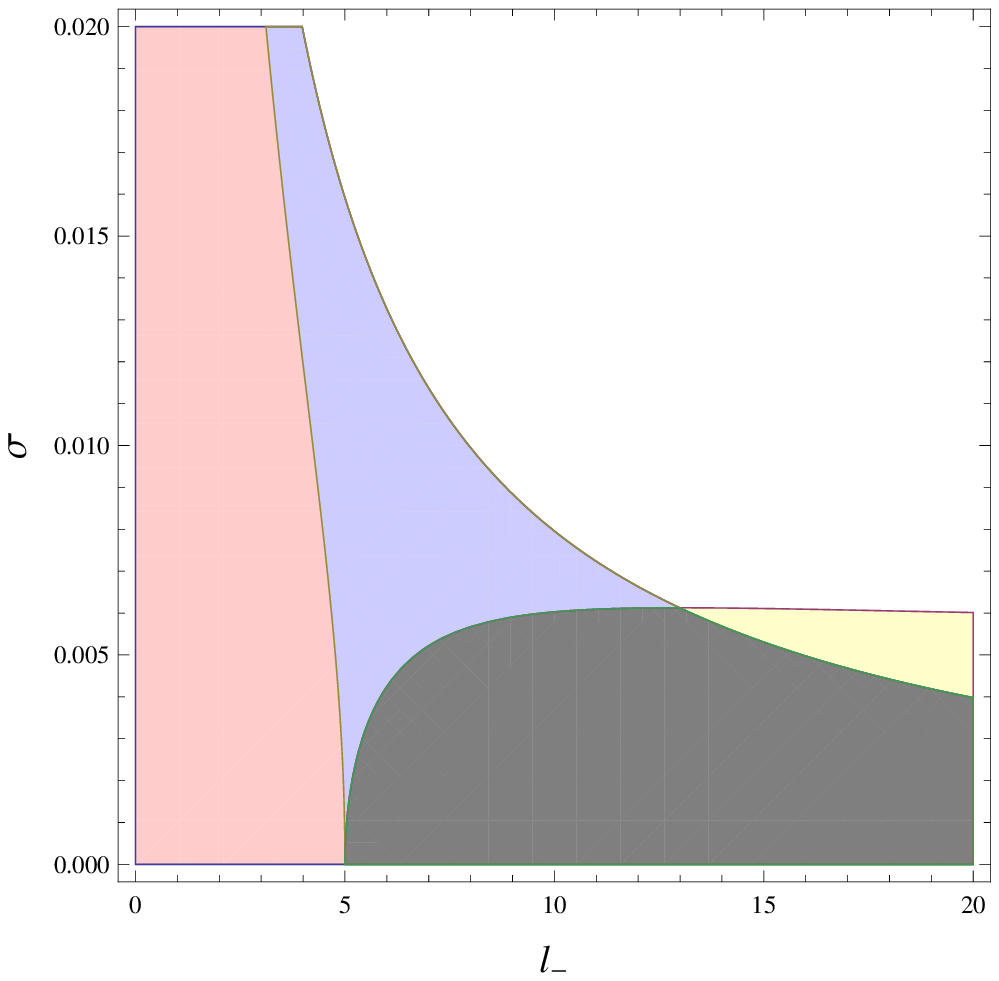}
\includegraphics[scale=0.54]{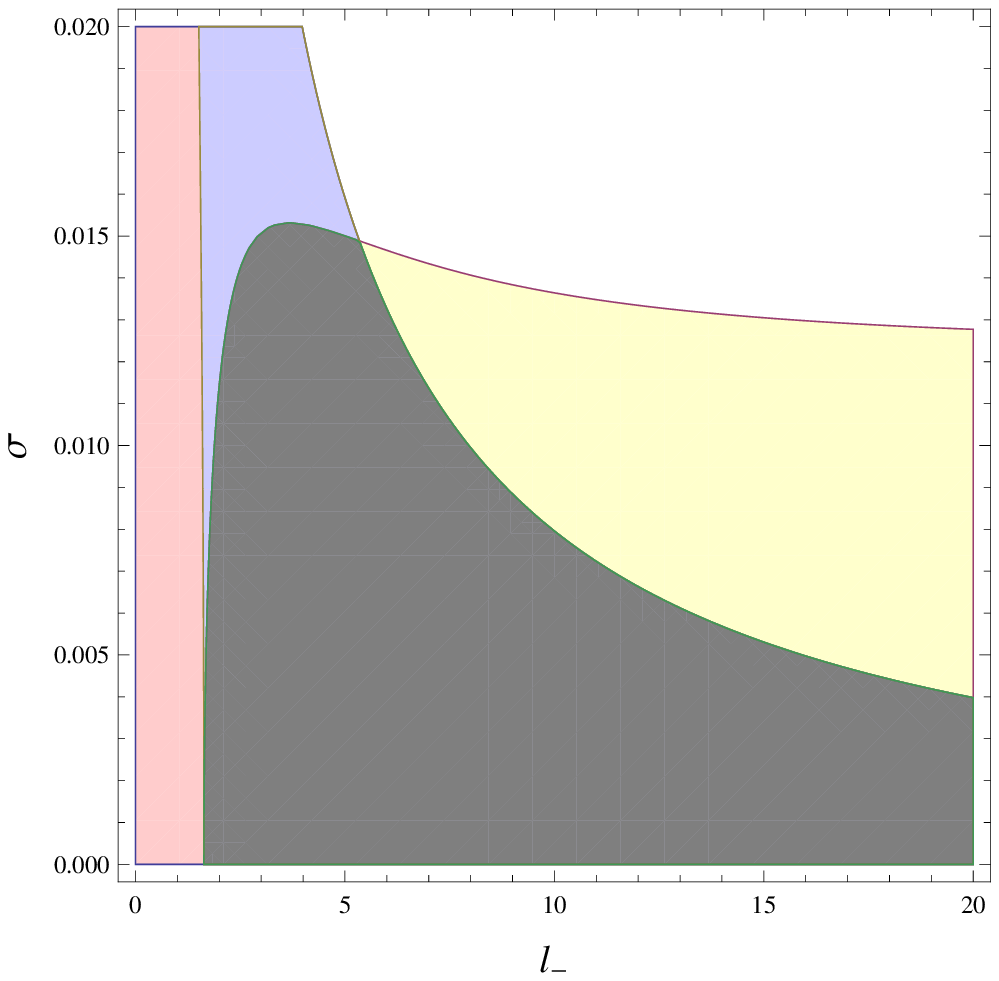}
\caption{\label{fig:parameters}Parameter spaces of $\ell_{-}$ and $\sigma$ by varying $M_{+}$ and $M_{-}$ (left: $M_{+}=1, M_{-}=5$, middle: $M_{+}=2.5, M_{-}=5$, right: $M_{+}=1, M_{-}=2.5$) that satisfy Eq.~(\ref{eq:cond1}) (red), Eq.~(\ref{eq:cond2}) (yellow), and Eq.~(\ref{eq:cond3}) (blue). The overlapped region is denoted by the black colored region.}
\end{center}
\end{figure}

\subsection{Parameter spaces and probability}

First, we need to check whether the conditions for the tunneling from the past horizon are well satisfied or not. The required conditions are as follows: (1) there exists two zeros $r_{1,2}$ satisfying $V(r_{1,2}) = 0$ and extrinsic curvatures should satisfy the conditions (2) $\beta_{\pm}(r_{1}) < 0$ and (3) $\beta_{\pm}(r_{2}) > 0$ in order to make sure that the shells are initially located beyond the Einstein-Rosen bridge and finally appears at the right side of the Penrose diagram.

\begin{figure}
\begin{center}
\includegraphics[scale=0.8]{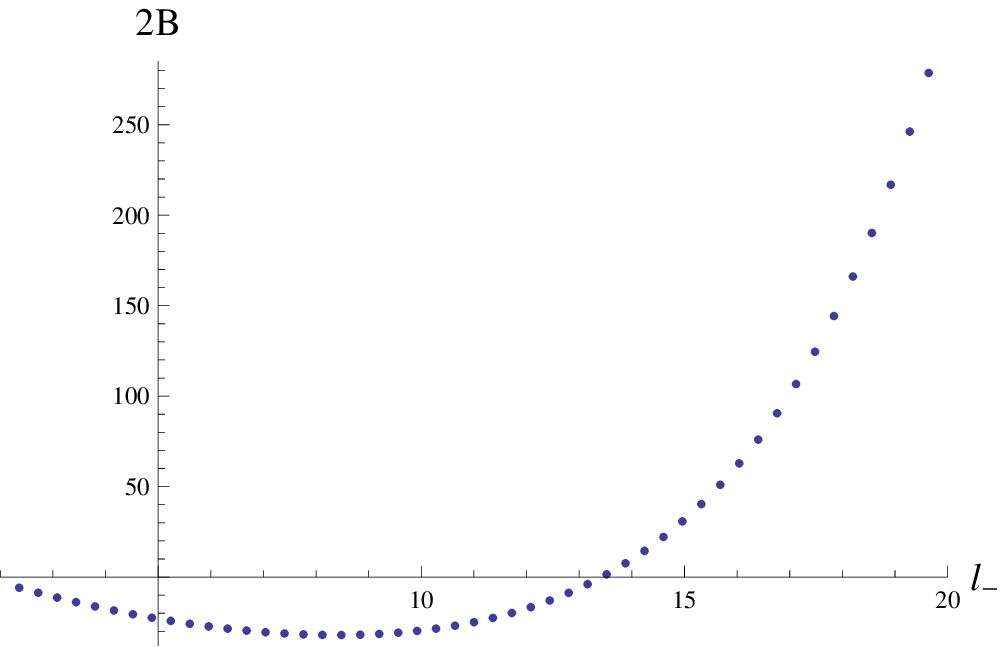}
\includegraphics[scale=0.8]{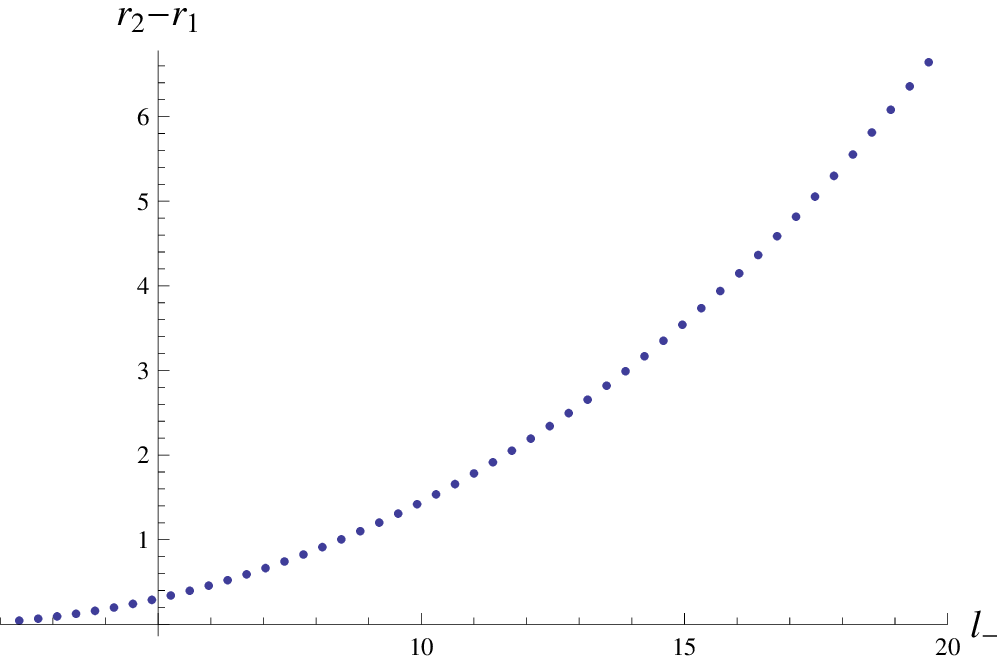}
\caption{\label{fig:1}$2B$ and $r_{2} - r_{1}$ by varying $\ell_{-}$ for a given $M_{+} = 1$, $M_{-} = 2.5$, and $\sigma = 0.001$. There exists an exponentially preferred region ($2B < 0$), while for the large $\ell_{-}$ limit, the probability is exponentially suppressed.}
\includegraphics[scale=0.8]{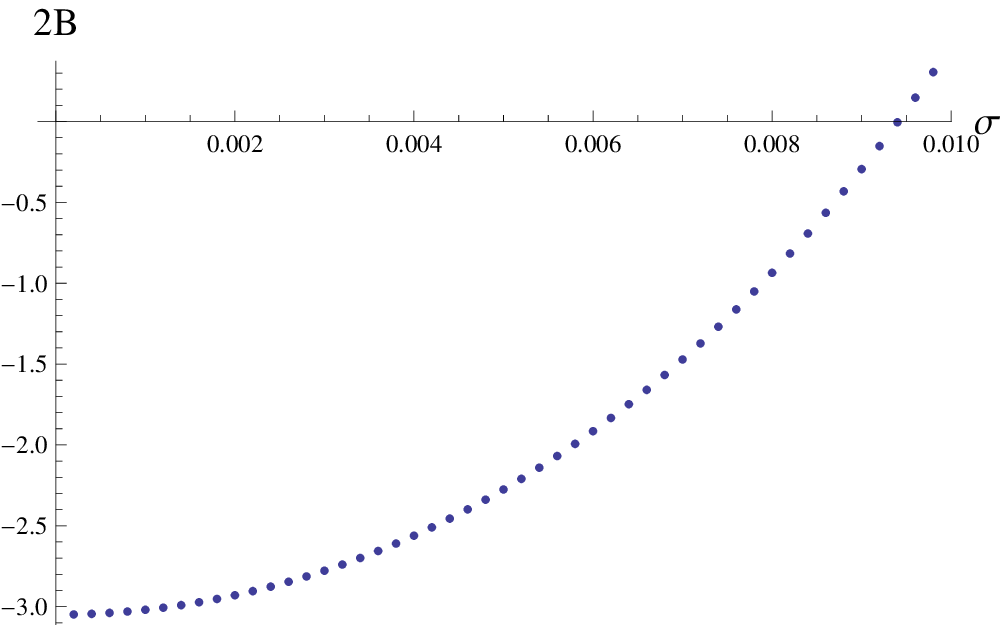}
\includegraphics[scale=0.8]{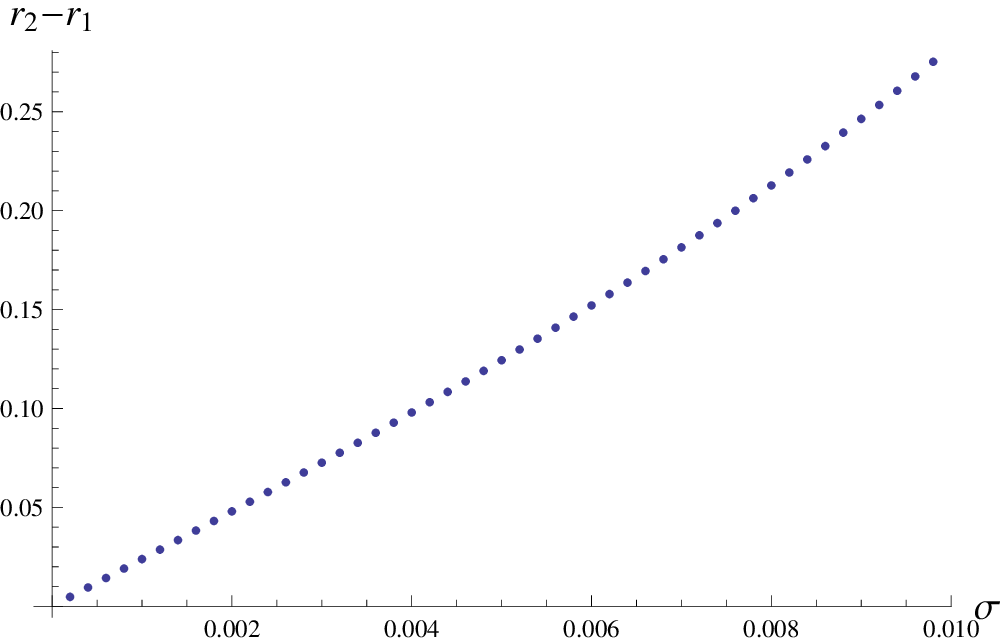}
\caption{\label{fig:2}$2B$ and $r_{2} - r_{1}$ by varying $\sigma$ for a given $M_{+} = 1$, $M_{-} = 2.5$, and $\ell_{-} = 2$.}
\end{center}
\end{figure}

Keeping Eq.~(\ref{eq:beta}) in mind, $r_{1}$ has a negative extrinsic curvature, if
\begin{eqnarray}
r_{1}^{3} < \frac{2\omega}{\ell_{-}^{-2} + 16\pi^{2} \sigma^{2}} \equiv r_{1*}^{3}
\end{eqnarray}
is satisfied and $r_{2}$ has a positive extrinsic curvature, if
\begin{eqnarray}
r_{2}^{3} > \frac{2\omega}{\ell_{-}^{-2} - 16\pi^{2} \sigma^{2}} \equiv r_{2*}^{3}
\end{eqnarray}
is satisfied. The later equation assumes the relation $\ell_{-} < 1/4\pi\sigma$. Now let us impose the following conditions: $V(r_{1*}) > 0$ and $V(r_{2*}) > 0$. If these conditions are satisfied, then automatically, there exist two zeros of $V(r)$, since $V(r)$ goes to negative infinity for small and large $r$ limits and we can also check that there should be two zeros if there exists a point that satisfies $V(r) > 0$. In addition to this, $r_{1} < r_{1*}$ and $r_{2} > r_{2*}$ must be automatically satisfied.

Therefore, it is enough to check the following three relations:
\begin{eqnarray}
\ell_{-} &<& 1/4\pi\sigma, \label{eq:cond1}\\
V(r_{1*}) &>& 0, \label{eq:cond2}\\
V(r_{2*}) &>& 0. \label{eq:cond3}
\end{eqnarray}
Then one can define a tunneling from the past horizon. The allowed parameter spaces for several combinations of $M_{+}$ and $M_{-}$ are plotted in Fig.~\ref{fig:parameters}.

Based on these allowed parameter spaces, we can calculate the tunneling probabilities.
\begin{itemize}
\item[--] By fixing $M_{\pm}$ and $\sigma$, we observe $2B$ by varying $\ell_{-}$ (Fig.~\ref{fig:1}). For the small $\ell_{-}$ limit, there appears an exponentially enhanced region ($2B < 0$). This is due to the fact that $\omega > 0$ and hence the difference of the areal entropy is always negative. However, as $\ell_{-}$ increases, $r_{2} - r_{1}$ increases\footnote{There is a tendency that $r_{2} - r_{1}$ increases as $\ell_{-}^{-2} - 16\pi^{2} \sigma^{2}$ approaches to zero, since such a limit corresponds the condition that $r_{2} = \infty$.} and hence the shell dynamics requires the entropy cost more and more. Therefore, the probability is exponentially suppressed for the large $\ell_{-}$ limit.
\item[--] By fixing $M_{\pm}$ and $\ell_{-}$, we observe $2B$ by varying $\sigma$ (Fig.~\ref{fig:2}). The tendency is similar such that as $r_{2}-r_{1}$ increases, the probability decreases.
\end{itemize}

\subsection{Stability of the Einstein-Rosen bridge}

Therefore, if $2B < 0$, then the Einstein-Rosen bridge is no more stable. After the tunneling, it will evolve to a larger horizon and hence the communication between the left side and the right side of the Einstein-Rosen bridge will be impossible. However, it is no more clear whether we can assume a certain vacuum of a given eternal black hole or a given Einstein-Rosen bridge.

Basically, there can be two ways to understand this. First, such a tunneling from the past horizon is impossible by some unknown reasons. For example, one may say that the white hole region must be hidden by a star interior. Of course, as we have mentioned in introduction, the mathematical use of the Einstein-Rosen bridge and the white hole region is still important in the context of the ER=EPR conjecture or eternal black holes.

Then the other alternative is to assume that the conditions for our tunneling process are prohibited by parameters. For example, we need specific conditions for $M_{\pm}$, $\sigma$, and $\ell_{-}$. Especially, $\ell_{-}$ is the physical parameter that needs to be assumed by the underlying theory, while the other parameters can be relatively freely chosen by situations. As we have observed in the previous section, if $\ell_{-}$ is large enough, i.e., if the anti-de Sitter vacuum energy is very close to zero, then such a process will be exponentially suppressed and one can obtain a quasi-stable Einstein-Rosen bridge.

In addition to this, there are several speculative questions:
\begin{itemize}
\item[--] For astrophysical black holes, there is no physical Einstein-Rosen bridge and hence there might be no tunneling from the past horizon. However, if we trust the ER=EPR conjecture strongly, then Hawking radiation and interior of the black hole can be connected by an Einstein-Rosen bridge due to the entanglements. Then there appears a non-perturbative process between two parts because of the Einstein-Rosen bridge. Now the question is this: \textit{does the ER=EPR conjecture induce a new non-perturbative effect}? If this is possible, then such a tunneling from the past horizon is a good way to confirm or falsify the ER=EPR conjecture, since it was absent previously but now appears by assuming the conjecture.
\item[--] Since the tunneling is from the left side to the right side of the Einstein-Rosen bridge, one may ask as follows: \textit{does the tunneling carry information from left to right side}? In terms of causality, it seems impossible since the horizon should increase after the tunneling. However, if the shell can carry any information or can affect entanglements between the left side and the right side, then such a tunneling process will do an important role to study quantum field theory in the anti-de Sitter space.
\item[--] If we allow such a tunneling, then \textit{can we give a single boundary at infinity} for an eternal black hole? Rather, it would be better to say that the boundary should be a superposition of several different anti-de Sitter vacuum \cite{Yeom:2016qec}. Therefore, after we sum over all histories, the bulk geometry will look very different from our intuitive pictures.
\end{itemize}
These are interesting questions but we do not know the answer. We would like to leave them for the future research topics.

\section{\label{sec:con}Conclusion}

In this paper, we investigated a tunneling process such that the shell was initially inside the past horizon, i.e., it was in the other side of the Einstein-Rosen bridge, but later tunnels to outside the horizon. For several parameter spaces, we observed that such a process is exponentially preferred. Then what does this mean?

We may illustrate several possibilities. First, there may be a problem of our theoretical setup itself. For example, we have regarded the past horizon and the other side of the Einstein-Rosen bridge as physical realities, but they might be problematic. Of course, on the other point of view, we can advocate out setting by revoking the work of Hartle and Hawking \cite{Hartle:1976tp} and the ER=EPR conjecture \cite{Maldacena:2013xja}.

Second, one may interpret that our analysis is not wrong, but our setup requires a specific parameters that may not be realistic in our universe. If this is the case, then we can explain the reason why our black holes are stable. However, this may mean that the stability of the Einstein-Rosen bridge cannot be guaranteed in generic cases.

In this paper, we cannot conclude clearly, but we want to mention that there is a possibility such that the Einstein-Rosen bridge can cause the instability of the original black hole to a deeper vacuum. If this is true, then we need to care about several new possibilities in the context of the ER=EPR conjecture or the quantum field theory in anti-de Sitter. We have suggested several interesting questions and we leave them for future research projects.

\newpage

\section*{Appendix A. Derivations of several formulas}

Let us start from the Israel's junction equation:
\begin{eqnarray}
\epsilon_{-} \sqrt{\dot{r}^{2}+f_{-}(r)} - \epsilon_{+} \sqrt{\dot{r}^{2}+f_{+}(r)} = 4\pi r \sigma.
\end{eqnarray}
If both sides are squared, we obtain
\begin{eqnarray}
2\dot{r}^{2} + \left( f_{+} + f_{-} - 16 \pi^{2} r^{2} \sigma^{2} \right) = 2 \epsilon_{+} \epsilon_{-} \sqrt{\left( \dot{r}^{2} + f_{-} \right) \left( \dot{r}^{2} + f_{+} \right)}.
\end{eqnarray}
If both sides are squared once again, then one can subtract the $\dot{r}^{4}$ term. The result is
\begin{eqnarray}
\dot{r}^{2} + \frac{4f_{+}f_{-} - \left( f_{+} + f_{-} - 16 \pi^{2} r^{2} \sigma^{2} \right)^{2}}{64 \pi^{2} r^{2} \sigma^{2}} = 0,
\end{eqnarray}
where this is equivalent to Eq.~(\ref{eq:form2}). From this, the extrinsic curvatures are
\begin{eqnarray}
\beta_{\pm} &=& \frac{\left( 1 - 2M_{-}/r + r^{2}/\ell_{-}^{2} \right) - \left( 1 - 2M_{+}/r + r^{2}/\ell_{+}^{2} \right) \mp 16 \pi^{2} \sigma^{2} r^{2}}{8\pi \sigma r}\\
&=& \frac{2 \left(M_{+} - M_{-}\right)/r + \left( r^{2}/\ell_{-}^{2} - r^{2}/\ell_{+}^{2} \right) \mp 16 \pi^{2} \sigma^{2} r^{2}}{8\pi \sigma r}.
\end{eqnarray}
If we insert $M_{-} = M_{+} + \omega$ and $\ell_{+} = \infty$, we obtain Eq.~(\ref{eq:beta}).

\section*{Appendix B. More on the Euclidean action integration}

In this appendix, we discuss more details on the Euclidean action integration based on \cite{Gregory:2013hja}. In the thin-shell limit, the Euclidean action integration can be presented by
\begin{eqnarray}
2 S_{\mathrm{E}} = 2 \left( S_{\mathrm{E}}^{-} + S_{\mathrm{E}}^{+} + S_{\mathrm{E}}^{\mathrm{shell}} + S_{\mathrm{E}}^{\mathrm{cusp}} \right),
\end{eqnarray}
where we multiplied two since we assumed that the integration only covers the half period of the solution. The first and the second term of the right-hand side are the volume integration for inside and outside the shell, the third term is the action integration on the shell, and the last term is the contribution from the cusp of the solution (if it exists). After several steps of calculations, we obtain
\begin{eqnarray}
2 \left( S_{\mathrm{E}}^{-} + S_{\mathrm{E}}^{+} + S_{\mathrm{E}}^{\mathrm{cusp}} \right) = \beta M_{+} - \frac{\mathcal{A}}{4},
\end{eqnarray}
where $\beta = 8\pi M_{+}$ and $\mathcal{A}_{h}$ is the area of the horizon. One can derive this by a direct calculation, but in this appendix, we just mention that the volume integration represents nothing but the usual interpretation of thermodynamics: $2S_{\mathrm{E}} = F/T = E/T - S$, where $F$ is the free energy, $E$ is the energy, $T$ is the temperature, and $S$ is the entropy. If we subtract this integration to the background solution, we will obtain the $- \Delta \mathrm{A}_{h}/4$ term.

The remained contribution is the shell integration \cite{Gregory:2013hja}: this becomes
\begin{eqnarray}
2 S_{\mathrm{E}}^{\mathrm{shell}} = - \frac{1}{2} \int \sigma \sqrt{h} d^{3}x - \frac{1}{16 \pi} \int \left( f_{+}' \dot{T}_{+} - f_{-}' \dot{T}_{-} \right) \sqrt{h} d^{3} x,
\end{eqnarray}
where $h_{\mu\nu}$ is the induced metric on the shell. Note that $f_{+} \dot{T}_{+} - f_{-} \dot{T}_{-} = - 4\pi r \sigma$ and $\sqrt{h}d^{3}x = 4\pi r^{2}d\tau$. By using them, we can substitute $\sigma$ and finally we obtain Eq.~(\ref{eq:formula}).

\section*{Acknowledgment}

DY is supported by the Korea Ministry of Education, Science and Technology, Gyeongsangbuk-Do and Pohang City for Independent Junior Research Groups at the Asia Pacific Center for Theoretical Physics. SK is supported by IBS under the project code, IBS-R018-D1.

\end{document}